# Comments on "Bridging Service-Oriented Architecture and IEC 61499 for Flexibility and Interoperability"

Kleanthis Thramboulidis
Electrical and Computer Engineering, University of Patras, Greece.

*Abstract*— **In the paper by W. Dai *et al.* (*IEEE Trans. On Industrial Informatics*, vol. 11, no. 3, pp. 771-781, June 2015), a formal mapping between IEC 61499 and SOA is presented and a SOA-based execution environment architecture is described. In this letter, the proposed in the above paper mapping and the execution environment architecture are discussed and their potential for the exploitation is disputed.**

## I. Introduction

Authors in [1] claim that they present a formal model for the application of SOA in the distributed automation domain in order to achieve flexible automation systems. However, they only present what they call "formal mapping" between IEC 61499 Function Blocks and SOA. Based on the presented "formal mapping" they describe an execution environment and demonstrate the flexibility of the proposed approach by a scenario for dynamic reconfiguration.

In this letter the proposed in [1] approach is discussed in the context of both the SOA paradigm and the IEC 61499 Function Block model, and its potential for exploitation is disputed. Even the claim of authors that the proposed formal mapping is just "to investigate what SOA features can be achieved in IEC 61499" raises questions on the contribution of [1].

The remainder of this letter is organized as follows. Section II discusses the "formal mapping" presented in [1, Sec. IV], the SOA-based execution environment architecture [1, Sec. V] and the dynamic reconfiguration [1, Sec. VI]. Section III concludes this letter.

## II. Discussion

SOA was introduced as an approach to design a software system to provide services either to end-user applications or other services distributed in a network, via published and discoverable interfaces [2]. It focuses on service specification and allows the developer to freely select the language that will be used to implemented the service. Thus, it is a higher level of software specification compared to the object oriented or procedural paradigms. Both of these programming paradigms can be used for the implementation of services. Mapping rules of WSDL to various languages have been defined to allow the implementation of services with these languages. The presented in [1] mapping is not within this context.

Authors adopt the IEC 61499 standard [1, Ref. 4] instead of the widely used in industry IEC 61131 [1, Ref. 1], for reasons they present in the paper. The IEC 61499 standard defines the Function Block (FB) as the basic construct for the development of industrial automation systems. The FB is a kind of type with (a) a specific interface that captures the inputs and outputs of its instances, i.e., FBIs, in terms of events and data, and (b) a state machine, called ECC, to specify the dynamic behavior of its instances. The notation used to specify the interface is of lower level of abstraction compared to the one used in object oriented languages.

### A.  Formal mapping between IEC 61499 Function Blocks and SOA

Authors in [1, Sec. 1] admit that SOA has been introduced to facilitate the creation of distributed networked computer systems. They also argue in [1, Sec. III] that "a PLC program could be built based on invoking external service libraries if external communication latencies are minimal compared to execution time of function blocks." However, the "formal mapping" they describe consists of a set of so called formal definitions for mapping SOA principles to IEC 61499 in order to interconnect FBIs on the same device. These definitions are next used as a guideline for the implementation of an IEC 61499 service-based execution environment for a device.

Authors do not use the IEC 61499 FB model in order to implement services. They map and implement an IEC 61499 based design of the software control system to a SOA based execution environment. Thus, they consider the IEC 61499 model as a higher level of specification and use the SOA paradigm to integrate the FBIs of an application running on the same device. Among the various problems of the presented approach, we discriminate the following:
1. the IEC 61499 model is completely inappropriate for expressing a SOA based design,
2. the given set of definitions may not be considered as a formal definition of an infrastructure for the application of SOA in industrial automation, and,
3. these definitions result to a completely inefficient execution environment, as argued in the following.

Based on [1, Definition 4], FBIs are service providers since each input event of an FBI is considered as a provided service. This is in contrast to the Definition 2 based on which an atomic service is used to represent every basic FBT. It is interesting to note that all the atomic FBs of fig. 4 [1], appear to provide the same services, i.e., INIT and REQ, even though they are of different types. Moreover, application events appear to be captured as data, e.g., NextSend, PrevSend, etc.

According to [1, Definition 5] there is a service repository in every IEC 61499 resource for the FBIs to register their provided services, as shown in [1, Fig. 1]. This is performed by having each FBI to register the service definitions or service contracts, as claimed by authors. WSDL is used by authors to define service contracts; the SOAP protocol is used to implement the interactions among FBIs in the same processing node [1, Sec. V].

Atomic services are defined for every basic FB [1, Definition 2] such as the ones defined to perform logic operations such as AND, OR, XOR as well as for merging (E_MERGE) and delaying (E_DELAY) events. Based on the above definitions and the performance analysis of [1, Sec. VII] an average overhead of 0.8 msec is introduced, even for the invocation of a simple services, for the case of persistent connections and 4.8 msec for not persistent connections. This leads to huge intra-device communication latencies compared to execution time of function blocks.

*B. The Execution Environment*

Authors describe in [1, Sec. V] an execution environment for IEC 61499 claiming that this is based on the formal definitions defined in the same paper.

From the definition of dynamic services it is extracted that not only input events are mapped to services but also the EC state algorithms. Data services are also defined to access internal variables of the FB instance. Service endpoints are also used for EC state actions, EC algorithms and EC actions. All these are stored in the service repository that means that SOAP and XML overhead is introduced even in the ECC execution time. Moreover, services are registered to the repository for every constituent FBIs of composite FB; thus the overhead from service utilization is also introduced at the composite FB level for the integration of its constituent parts. The WS-discovery protocol is utilized for service discovery from the resource's repository. Even though the approach focus on distributed systems the relation of the resource repository with the device external one, that would probably be used to register device's exposed services is not discussed.

For the presented execution environment, authors assume that EC algorithms are normally written in IEC 61131 languages and mainly ST and LD. However, this raises the question of portability that was considered one of the main factors for the selection of 61499 instead of the 61131, which as claimed in [1] does not provide code portability among various PLC vendors. On the other side it is claimed that code portability is achieved for FB library elements due to the use of their XML-based representation. It should be noted that an XML based representation for IEC 61131 (PLCopen) is already in the market for a long time.

*C. Dynamic reconfiguration*

Dynamic reconfiguration at the device level, which is considered as one benefit of the proposed architecture, imposes string real time constraints and complex algorithms not shown in [1]. No indication of time requirements for the execution of the actions of [1, Table I] are given; thus, the claim that this reconfiguration procedure is performed "without stopping normal operation", is completely arbitrary.

The described in [1, Table I] case study includes actions for deleting and creating event and data connections. The creation of event connections among FBIs has to be related to the publish/discover based interaction on which the proposed architecture is based. The resource management model described in IEC 61499 to support the IDE in the deployment process is not consistent with the publish/discover model that authors have adopted for the construction of the formal model [1, Sec. IV]. For example, the management command of IEC 61499 *"CREATE event connection"* expresses a different model from the publish/discover pattern. A coordinator, the IDE, enforces the construction of an event connection among the specific FBIs. This is not consistent with the publish/discover pattern and the authors' claim, according to which when an FBI "intends to invoke a particular logic from a service provider, the requested service will be located by the service repository for the service requester." Based on this, authors claim that "the service requester can access the service provider via sending messages"

An execution environment for IEC 16499 that supports run time reconfiguration with detailed performance measurements is presented in [3]. Based on this: a) the average value of the FB instance creation time is 20 $\mu$s, and b) the creation of an event connection has an average time of 1.87 $\mu$s, while its deletion has an average value of 1.8 $\mu$s, both with a standard deviation of about 0.5 $\mu$s. The publish/subscribe communication pattern of RTNet is used as a communication mechanism instead of web services and SOAP which introduce a huge overhead.

III. CONCLUSION

SOA has been evaluated by several research groups for its potential application in industrial automation systems. Research projects have resulted in the development of protocol stacks for the device level to allow the interconnection of the control PLCs with the upper layers of the manufacturing pyramid. However, SOAP and Web Services even though introduced in some PLCs have considerable performance overhead that is a big barrier in their use. Other technologies, such as IoT and the REST architecture, provide feasible solutions to this level of integration.

The use of SOAP, WSDL and WS-discovery protocol for the integration of the components of the controlling software of a device, but also for the implementation of composite FBs, as proposed in [1], greatly increases the performance overhead as well as the complexity at this level and is considered as completely unorthodox approach for utilizing SOA. Other technologies provide feasible solutions to this level of integration. SOAP has been developed to interconnect functionalities expressed in terms of software developed on heterogeneous hardware and/or software platforms, which are distributed over the internet. These two requirements, i.e., distribution and heterogeneity, do not exist in the single device IEC 61499 execution environment thus the cost of performance overhead and the complexity that its adoption introduces is without a benefit. An extensive discussion on the approach presented in [1] can be found in [4].